\documentclass{emulateapj}
\begin{document}

\title{Kinematics and excitation of the ram pressure stripped ionized gas filaments 
in the Coma cluster of galaxies\altaffilmark{1}}

\author{
Michitoshi Yoshida\altaffilmark{2},
Masafumi Yagi\altaffilmark{3},
Yutaka Komiyama\altaffilmark{3},
Hisanori Furusawa\altaffilmark{4},
Nobunari Kashikawa\altaffilmark{3},
Takashi Hattori\altaffilmark{5}
and
Sadanori Okamura\altaffilmark{6}
}

%\vspace {3cm}

\altaffiltext{1}{Based on data collected at the Subaru Telescope, which is operated by the
National Astronomical Observatory of Japan.}

\altaffiltext{2}{Hiroshima Astrophysical Science Center, Hiroshima University,
Higashi-Hiroshima, Hiroshima 739-8526, Japan; yoshidam@hiroshima-u.ac.jp}

\altaffiltext{3}{Optical and Infrared Astronomy Division, National 
Astronomical Observatory,
Mitaka, Tokyo 181-8588, Japan.}

\altaffiltext{4}{Astronomical Data Center, 
National Astronomical Observatory of Japan, Mitaka
Tokyo 181-8588, Japan.}

\altaffiltext{5}{Subaru Telescope, National Astronomical Observatory of 
Japan, 650 North
A'Ohoku Place, Hilo, HI 96720, USA.}

\altaffiltext{6}{Department of Astronomy, University of Tokyo,
Tokyo 113-0033, Japan.}

%------------------------------------------------------------------------------

\begin{abstract}

We present the results of deep imaging and spectroscopic observations of
very extended ionized gas (EIG) around four member galaxies of the Coma
cluster of galaxies: RB199, IC~4040, GMP~2923 and GMP~3071.
The EIGs were serendipitously found in an H$\alpha$ narrow band imaging
survey of the central region of the Coma cluster.
The relative radial velocities of the EIGs with respect to the systemic
velocities of the parent galaxies from which they emanate increase
almost monotonically with the distance from the nucleus of the
respective galaxies, reaching $\sim -400 - -800$ km~s$^{-1}$ at
around $40 - 80$ kpc from the galaxies.
The one-sided morphologies and the velocity fields of the EIGs are 
consistent with the predictions of numerical simulations of ram pressure 
stripping.
We found a very low-velocity filament ($v_{\rm rel} \sim -1300$ km~s$^{-1}$)
at the southeastern edge of the disk of IC~4040.
Some bright compact knots in the EIGs of RB199 and IC~4040 exhibit blue
continuum and strong H$\alpha$ emission.
The equivalent widths of the H$\alpha$ emission exceed 200 \AA\, 
and are greater than 1000 \AA\ for some knots.
The emission line intensity ratios of the knots are basically consistent
with those of sub-solar abundance \ion{H}{2} regions.
These facts indicate that intensive star formation occurs in the knots.
%The luminosity and the size of the knots are consistent with the
%scenario that the knots eventually evolve into ultra compact dwarfs.
Some filaments, including the low velocity filament of the IC~4040
EIG, exhibit shock-like emission line spectra, suggesting that shock 
heating plays an important role in ionization and excitation of the EIGs.

\end{abstract}

%-------------------------------------------------------------------------------


%-------------------------------------------------------------------------------

\section{Introduction}

Clusters of galaxies are ideal laboratories for investigating environmental
effects in galaxy evolution. 
The extreme environments of the clusters --- the high number density 
of galaxies, strong gravitational field, and dense hot gas filling 
inter-galactic space --- allow us to study how these environmental factors
affect the evolution of cluster member galaxies.
A substantial body of observational evidence suggests that cluster
environment significantly affects cluster galaxy evolution; for example, 
the morphology - density relation \citep{Dressler1980,Postman1984,Dressler1994,
Wel2007}, the redshift - blue galaxy fraction relation \citep{Butcher1978,Butcher1984}, 
or the density - H~{\sc i} deficiency relation \citep{Cayatte1990,Cayatte1994,
Solanes2001}.
Researchers have proposed various physical mechanisms potentially responsible
for these observational facts \citep{Boselli2006}, including galaxy-galaxy 
interaction \citep{Byrd1990}, galaxy harassment \citep{Moore1996}, tidal interaction 
between galaxies and cluster potential \citep{Byrd1990,Henriksen1996}, 
galaxy starvation \citep{Bekki2003}, and ram pressure stripping \citep[hereafter referred as to RPS]{Gunn1972}.

Although it is not clear which mechanism plays the most important role
in the morphological and color evolution of cluster galaxies, it is at least 
certain that rapid removal of the interstellar gas of galaxies
is a key factor \citep{Okamoto2003,vanGorkom2004,Boselli2008}. 
In particular, RPS
caused by fast interaction between galaxies and the hot intra-cluster medium (ICM)
must play a very important role in gas removal \citep{Vollmer2001b,Fujita2004,Cortese2011}.
A number of observational studies have focused on this subject.
The H~{\sc i} gas of galaxies is systematically deficient in the cores of nearby
rich clusters \citep{Cayatte1990,Cayatte1994,Bravo2000,Bravo2001}.
Deep H~{\sc i} imaging revealed one-sided H~{\sc i} gas flows around 
late-type galaxies in the Virgo cluster \citep{Oosterloo2005,Chung2007,Chung2009}.
Asymmetric distributions of molecular gas found in some cluster galaxies also 
suggest that the molecular gas is affected by ram pressure from the hot ICM 
\citep{Vollmer2001a,Vollmer2003,Vollmer2008,Vollmer2009b,Sivanandam2010}.
Recently, elongated X-ray gas-flows from galaxies have been found in nearby
clusters \citep{Sun2007,Sun2010,Randall2008}.
In addition, many case studies have focused on ionized gas flows from cluster member
galaxies probed by H$\alpha$ emission \citep{Kenney1999,Gavazzi2001,Gavazzi2003,
Yoshida2002,Yoshida2004,Sakai2002,Chemin2005,Cortese2006,Kenney2008}.
Some pieces of evidence of dust stripping in the Virgo galaxies have also been reported 
\citep{Crowl2005,Abramson2011}.

We serendipitously discovered a number of very extended ionized gas filaments
(EIG) around the member galaxies of the Coma cluster
(\citeauthor{Yagi2007} \citeyear{Yagi2007}; 
\citeauthor{Yoshida2008} \citeyear[hereafter YY08]{Yoshida2008};
\citeauthor{Yagi2010} \citeyear[hereafter YY10]{Yagi2010}).
The filaments are extended toward one side of the parent galaxies (YY10).
The most remarkable example of the one-sided extension is the 60-kpc straight
H$\alpha$ filament found around D100 \citep{Yagi2007}.
Another interesting object is the complex of ionized gas filaments, star-forming
blue knots and young stellar filaments extended from an E+A galaxy RB199 
(``fireballs''; YY08).
The morphology of the latter very much resembles strings of compact blue knots
around the two galaxies falling into $z\approx0.2$ clusters found 
by \citet{Cortese2007}.
\citet{Cortese2007} interpreted these peculiar features as star-forming regions 
in the gas stripped from the parent galaxies by the combined action of tidal
interaction and RPS.
The fireballs is a low-redshift counterpart of these features (YY08).
Recently, \citet{Hester2010} reported a similar structure around a low surface
brightness galaxy IC~3418 in Virgo cluster (see also \citeauthor{Fumagalli2011}
\citeyear{Fumagalli2011}).

YY10 found some properties specific to the EIGs' parent galaxies.
The radial velocities of the parent galaxies are distributed
either at the blue or at the red edge of the velocity distribution
of Coma member galaxies.
In addition, almost all the parent galaxies have blue colors; about 60\%\ of
the parent galaxies are distributed well below (bluer side) from the 
color-magnitude relation (red sequence) of the early type galaxies of the 
Coma cluster.
These facts suggest that the parent galaxies are late-type galaxies 
which have recently fallen into the cluster.
They have not yet been well processed in the cluster environment. 
Further, YY10 classified the EIGs into three categories: 
(1) disk star-formation + connected H$\alpha$ gas, (2) no disk star-formation
with connected H$\alpha$ gas, and (3) detached H$\alpha$ gas.

\citet{Smith2010} found on $GALEX$ images one-sided UV tails
around a number of Coma member galaxies whose morphologies
resemble the EIGs.
More than half of their objects overlapped the EIG galaxies discovered
by YY10.
They assessed the statistical characteristics of these UV tail galaxies 
and concluded that these galaxies are newcomers to the Coma cluster
\citep{Smith2010}.
Good correlation between the EIGs and the UV tails suggests that these 
two kinds of features were created by the same mechanism; the most 
plausible one is RPS.
Furthermore, this indicates active star formation occurs in the stripped 
gas in the cluster environment.

Here we present the results of optical spectroscopic observations of the 
EIGs around four Coma member galaxies.
We assumed that the cosmological parameters 
($h_0$, $\Omega_m$, $\Omega_\lambda$) = (0.73, 0.24, 0.72) and 
the distance modulus of the Coma cluster is 35.05 \citep{Yagi2007}.
The linear scale at the Coma cluster is 474 pc arcsec$^{-1}$ under this 
assumption.

\section{Observation and Data Reduction}

Deep spectroscopic observations of four Coma member galaxies, RB199, 
IC~4040 (GMP~2559), GMP~2923 and GMP~3071 was performed using
the faint object spectrograph FOCAS \citep{Kashikawa2002} attached to
the Subaru Telescope on 20 May 2009.
Two multi-slit masks for each of RB199 and IC~4040 were used to cover
the complicated structure of the emission
line regions over as wide an area as possible.
GMP~2923 and GMP~3071 were observed using one multi-slit mask.
The positions and lengths of the slits are shown in Figures 1 -- 4.
The width of each small slit was 0\arcsec.8.
The lengths of the slits ranged from 10\arcsec\ to 30\arcsec.
A 300 grooves mm$^{-1}$ grism, whose straight light travel wavelength
is 6500 \AA, was used.
The full width at half maximum (FWHM) of the sky emission
lines around 6700 \AA\ is 7.8 \AA, which means that the spectral 
resolving power $R$ of our spectroscopy is 850 (the velocity resolution
is about 350 km~s$^{-1}$) over H$\alpha$ region.
Table 1 presents the observation log.

\begin{figure}[htb]
\begin{center}
\includegraphics[scale=1.3]{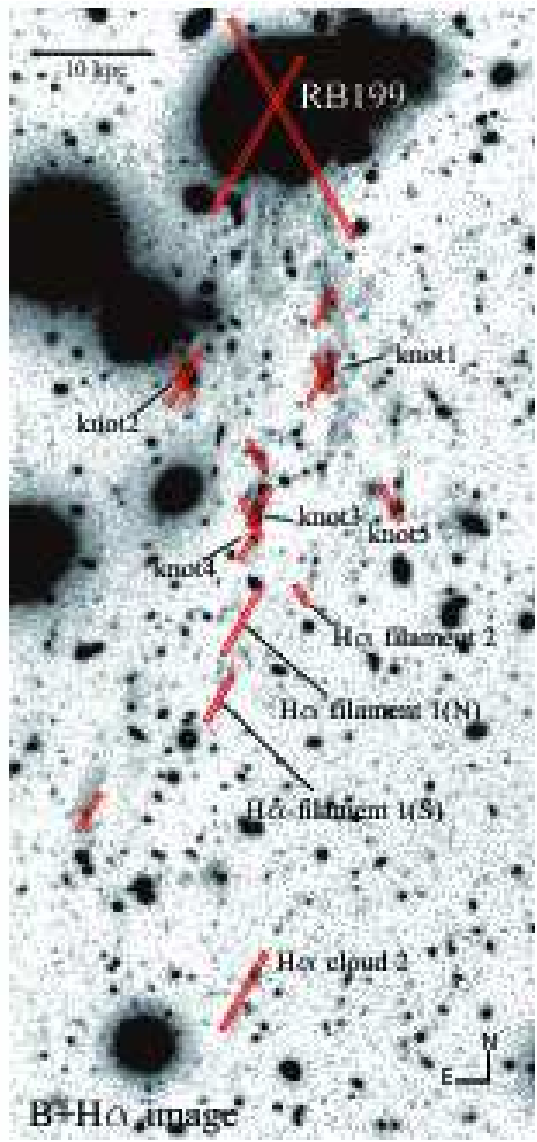}
\end{center}
\caption{
\label{RB199slits} Slit positions of the spectroscopy of the extended ionized gas (EIG) of RB199
(``fireballs'': YY08) overlaid on a composite image of the H$\alpha$ narrow band 
and $B$ band images. All the images used in Figures 1 -- 4 were
taken with Suprime Cam attached to Subaru Telescope. Details of the observations of these images
were described previously in YY10.
}
\end{figure}

\begin{figure}[htbp]
\begin{center}
\includegraphics[scale=1.15]{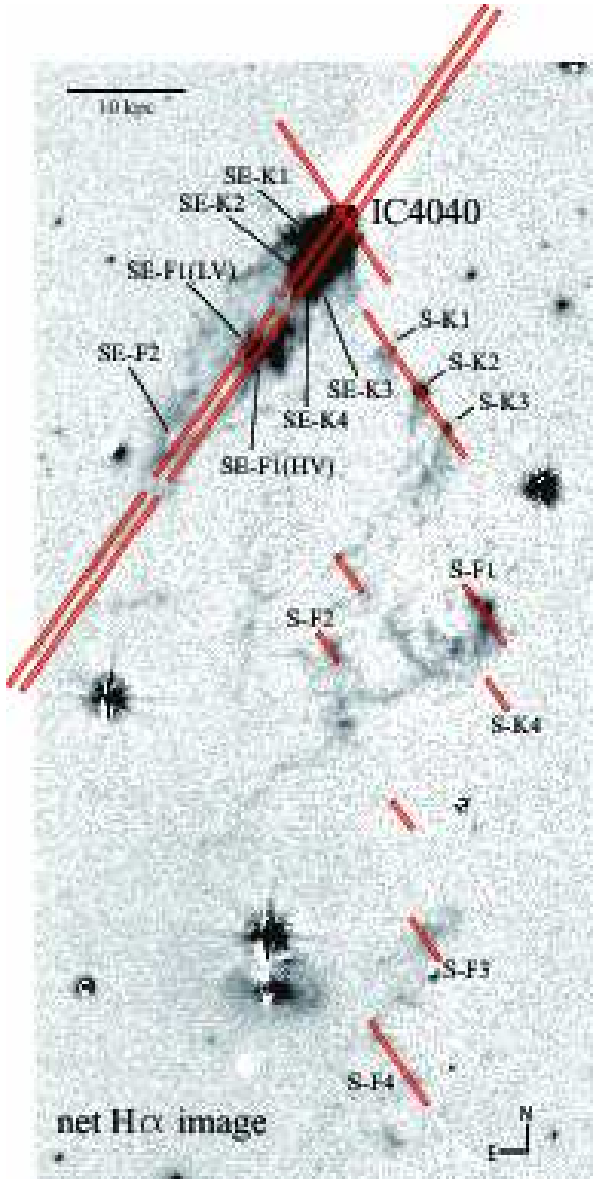}
\end{center}
\caption{
\label{IC4040slits} Slit positions of the spectroscopy of the EIG of IC~4040 overlaid on the net H$\alpha$
image.
}
\end{figure}

\begin{figure}[htbp]
\begin{center}
\includegraphics[scale=0.8]{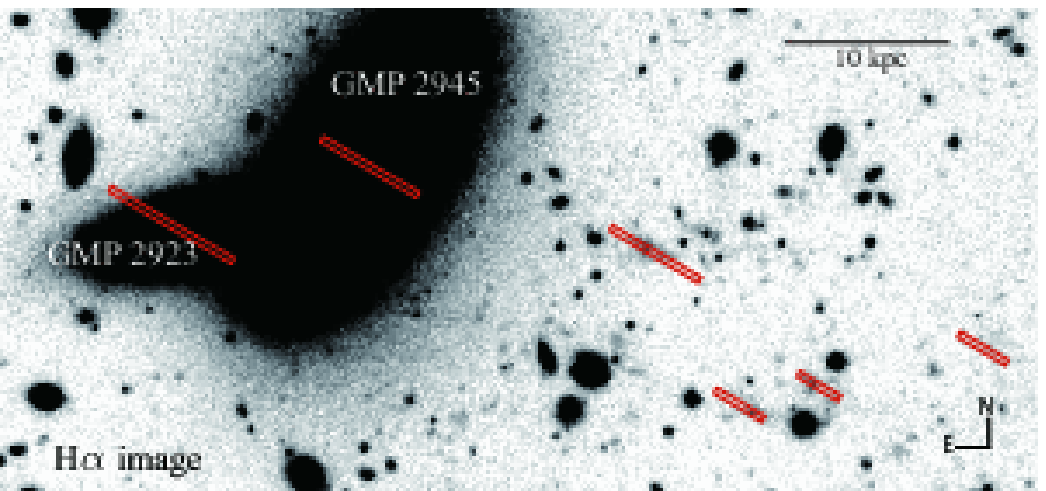}
\end{center}
\caption{
\label{GMP2923slits} Slit positions of the spectroscopy of the EIG of GMP~2923 overlaid on the H$\alpha$
narrow band image.
}
\end{figure}

\begin{figure}
\begin{center}
\includegraphics[scale=0.8]{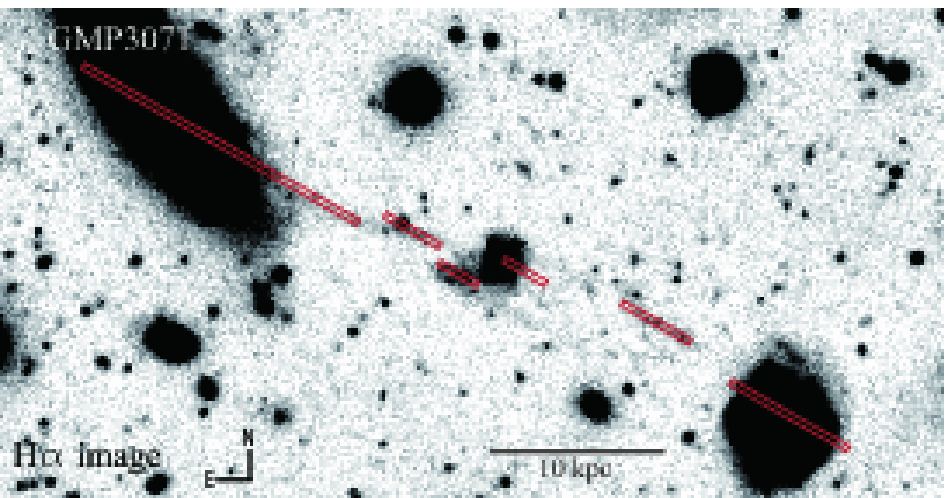}
\end{center}
\caption{
\label{GMP3071slits} Slit positions of the spectroscopy of the EIG of GMP~3071 overlaid on the H$\alpha$
narrow band image.
}
\end{figure}

\begin{table*}[tbp]
\caption{Spectroscopy Observation Log}
\begin{center}
\begin{tabular}{lcccc}
\hline
mask ID & object & center & PA & exposure \\
\hline
MS0631 & IC~4040 & 13$^h$00$^m$43.2$^s$ +28$^\circ$03\arcmin 06\arcsec & 215$^\circ$ & 6$\times$600sec  \\
MS0632 & IC~4040 & 13$^h$00$^m$38.4$^s$ +28$^\circ$04\arcmin 42\arcsec & 125$^\circ$ & 3$\times$300sec $+$ 3$\times$600sec  \\
MS0632-off & IC~4040 & 13$^h$00$^m$38.6$^s$ +28$^\circ$04\arcmin 45\arcsec & 125$^\circ$ & 1$\times$600sec  \\
MS0634 & RB199   & 12$^h$58$^m$43.3$^s$ +27$^\circ$46\arcmin 19\arcsec & 150$^\circ$ & 6$\times$600sec  \\
MS0635 & RB199   & 12$^h$58$^m$38.7$^s$ +27$^\circ$44\arcmin 35\arcsec & 210$^\circ$ & 6$\times$600sec  \\
MS0636 & GMP2923,3071 & 12$^h$59$^m$58.4$^s$ +27$^\circ$44\arcmin 58\arcsec & 60$^\circ$ & 2$\times$600sec \\
\hline
\end{tabular}
\end{center}
\end{table*}

Data reduction was performed in standard manner.
The bias level of each frame was estimated with average count of the 
over-scan region of the CCD and then subtracted from each frame.
Flat-fielding was done using dome flat frames. 
Wavelength calibration was performed using sky emission lines.
The rms residual of wavelength fitting for sky lines was 0.14 \AA\ in 
the wavelength region of 6000 \AA\ - 7000 \AA. 
Because there are no strong sky emission lines blueward of 5500 \AA, 
wavelength calibration is not accurate as in the longer wavelength region.
The rms error was 0.3 \AA\ at 5300 \AA.
Thus, we measured the radial velocities of the EIGs using the H$\alpha$ line only.

Flux calibration was initially conducted using spectro-photometric standard star data.
However, the standard star data were taken with an order sorting filter (Y47) 
whose cut-off wavelength was 4700 \AA,
while the filter was not used for the observation of the EIGs.
Thus, we had to correct the transmission curve of Y47 to ensure accurate
flux calibration for the wavelength region blueward of 4700 \AA.
We divided the initial flux-calibrated frames by the transmission curve of 
Y47, which was normalized to 1 at the maximum transmission wavelength
(6000 \AA).
We checked validity of the above recalibration procedure by comparing the 
calibrated spectrum of an elliptical galaxy (SDSS J125845.26+274655.0) 
taken with one slit of the mask 
MS0635 with the spectra of eight early type galaxies listed in \citet{Kennicutt1998}.
The type of the reference galaxies ranges from E0 to S0.
All the spectra were normalized at 6000 \AA. 
The rms relative difference between the spectrum of our E galaxy and the
spectrum of each reference early type galaxy was less than 7\% for the
wavelength region between 4300 \AA\ and 7000 \AA.
Thus we adopted that the recalibration we made was correct enough for
further analysis.
The signal-to-noise ratio of the measured spectra shorter than 4300 \AA\ 
was too low to enable confident comparison, so we do not include spectra 
shorter than 4300 \AA\ in the following analysis and
discussion.

We fitted gaussian functions to the emission lines to measure the fluxes and
the velocities of the lines.
Although it was difficult to measure velocity dispersion accurately because of low
spectral resolution of our spectra, we obtained rough estimation of the velocity
dispersion $\sigma$ by applying a simple decomposition scheme; 
$\sigma = \sqrt{\sigma_{\rm obs}^2 - \sigma_{\rm sky}^2}$,
where ${\sigma_{\rm obs}}$ and ${\sigma_{\rm sky}}$ are
the dispersions of the gaussian functions fitted to the observed emission line
and that of the sky line, respectively
($f(\lambda) \propto {\rm exp}(-\lambda^2 / 2 \sigma^2)$; $\sigma = {\rm FWHM} / 2 \sqrt{2 {\rm ln} 2}$). 
One spectral resolution unit was sampled with eight pixels of the CCD.
This over-sampling enabled us to measure the emission line widths with
an accuracy at 10\%\ of the FWHM of the sky lines.
Since the FWHM of the sky lines was 7.8 \AA, we resolved the emission lines whose
FWHMs were wider than 8.6 \AA, which means that the lower limit of $\sigma$ we could
measure was $\approx 70$ km~s$^{-1}$ at 6700 \AA.

Imaging data of the target galaxies in deep narrow-band H$\alpha$ and
broad-band $B$ and $R_{\rm C}$ bands used in this paper are those
presented in YY10 for morphological study of the EIGs around these
galaxies. They were taken using Suprime Cam \citep{Miyazaki2002}
attached to the Subaru Telescope. The data were collected over three
observing runs from 2006 - 2009. See YY10 for the details of the
imaging observations and data reduction.

\section{Results}

\subsection{The morphologies and kinematics of the EIGs}

The overall morphological characteristics of the EIGs of our
target galaxies were previously reported by YY10. Here, we summarize
the details of the morphologies and give an interpretation to
them based on the newly obtained kinematical information.

\subsubsection{RB199}

The $B$, $R_{\rm C}$ and H$\alpha$ images of RB199 (fireball) were shown
in our previous work (YY08). 
The characteristics of the EIG of RB199 can be summarized as follows:
(1) The EIG consists of a number of bright star-forming knots, elongated 
blue filaments and more extended H$\alpha$ emitting gas.
Total length of the EIG is about 80 kpc.
(2) The absolute magnitudes of the bright knots range from $-12$ $-$ $-13$
mag, which are comparable to nearby dwarf galaxies.
(3) The bright knots emit both blue continuum light and H$\alpha$ emission.
The H$\alpha$ emitting regions are offset from the peak of the blue continuum emission.
(4) The blue filaments are connected to the main body of the galaxy, while
the H$\alpha$ filaments are connected to the tips of the blue filaments.
(5) The bright knots and blue filaments are detected in UV light by the $GALEX$
satellite.

\begin{figure}[htbp]
\begin{center}
\includegraphics[scale=0.33]{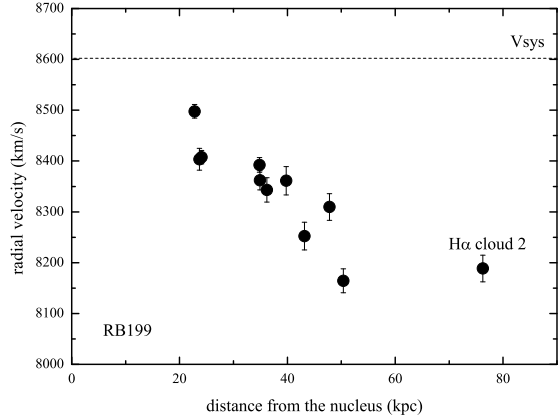}
\end{center}
\caption{
\label{RB199vel} The velocity field of the EIG of RB199. The systemic velocity of the galaxy is
determined by the H$\alpha$ absorption line of the nucleus.
}
\end{figure}

Figure 5 shows the velocity field of the EIG of RB199.
Relative radial velocities of the EIG filaments with respect to
the systemic velocity of RB199 increase with the distance from the
nucleus at almost a constant rate of $-7$ km~s$^{-1}$~kpc$^{-1}$.
The relative radial velocity of the most distant
cloud of the EIG, H$\alpha$ cloud 2 (YY08), is roughly consistent with
the extrapolation of the line of constant increase,
which suggests that this cloud is really a part of the EIG.

\begin{figure}
\begin{center}
\includegraphics[scale=0.33]{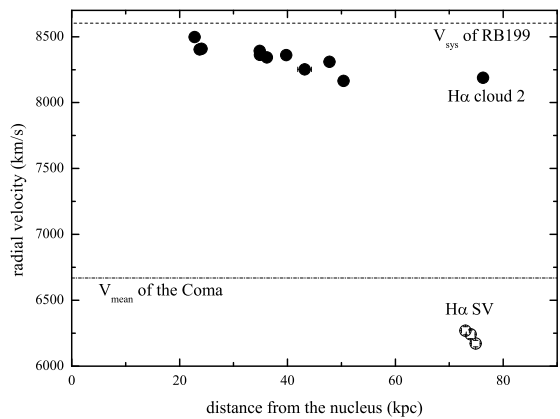}
\end{center}
\caption{
\label{RBvel-all} The velocity fields of the EIG of RB199 and the low velocity filaments 
overlapping H$\alpha$ filament 2. 
}
\end{figure}

We found that a faint H$\alpha$ component with a velocity of
$\approx 6250$ km~s$^{-1}$  overlaps along the line of sight to 
H$\alpha$ cloud 2.
Figure 6 shows the velocity field of the RB199 EIG with this slow 
velocity (SV) component.
The velocity of this SV component is slightly lower than that of the
mean velocity ($\approx 6900$ km~s$^{-1}$) of the Coma cluster.
The velocity difference between the SV component and H$\alpha$ cloud 2
is $\sim 2000$ km~s$^{-1}$, suggesting that the SV component is not 
associated with the RB199 EIG.

\subsubsection{IC~4040}

Figure 7 presents a pseudo-color image (blue, green and red colors represent
$B$, $R_{\rm C}$, and H$\alpha$ images, respectively) of the spiral galaxy 
IC~4040 taken with Subaru Suprime-Cam (YY10, reproduced by permission of the
American Astronomical Society; AAS).
The figure clearly shows a very extended EIG expelled from the galaxy disk.

\begin{figure}[htbp]
\begin{center}
\includegraphics[scale=0.95]{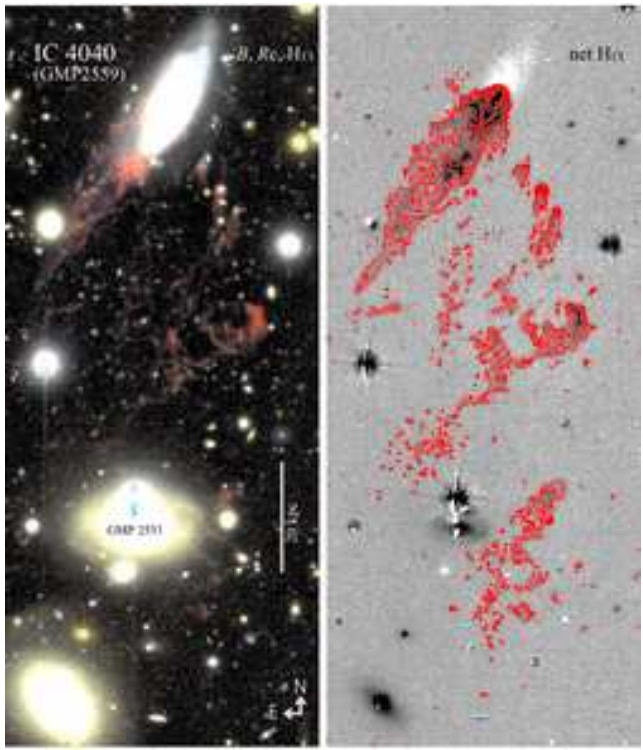}
\end{center}
\caption{
\label{IC4040color} A pseudo color image ($B$, $R_{\rm C}$ and H$\alpha$ composite: left panel) and 
net H$\alpha$ image (right panel) of IC~4040.
This figure is reproduced from YY10 by permission of the AAS.
}
\end{figure}

The structure of the EIG is very complicated.
Two major streams extend toward the southeast and south from the galaxy.
The southeast stream of the EIG is elongated almost linearly along the major
axis of the galaxy and its root is connected to the nuclear star-forming region. 
This southeast stream reaches $\sim 50$ kpc from the nucleus.
Figure 8 shows the detailed structure of the central bright part of the EIG.
The nucleus of the galaxy is surrounded by several bright star-forming knots.
This complex of the star-forming regions is truncated at the northwestern 
side of the nucleus.
The shape of the ionized gas distribution indicates a large-scale bow shock
and a giant gas flow toward the southeast from the nucleus.
The opening angle of the gas flow is about 65$^{\circ}$ and almost symmetric
with respect to the major axis of the galaxy.
Several bright H$\alpha$ knots (NW-K1, SE-K1, etc. see Figure 8)
surround the nucleus.
The brightest one, SE-K4, has a size of $r \sim 0.5$ kpc and is located at
7 kpc from the nucleus.
A large complex of diffuse H$\alpha$ clouds/filaments appears at around 
15 kpc from the nucleus.
Further away, a tangle of several filaments, with typical width of 1 kpc,
extends out to 50 kpc.
For convenience, we refer to the large complex and the filaments between 
20 kpc and 30 kpc as SE-F1 and SE-F2, respectively.
As described later, our spectroscopic observations revealed that the SE-F1
consists of at least
two kinematically-different components, a low-velocity (LV) component and
a high-velocity (HV) component.

\begin{figure}[htbp]
\begin{center}
\includegraphics[scale=0.95]{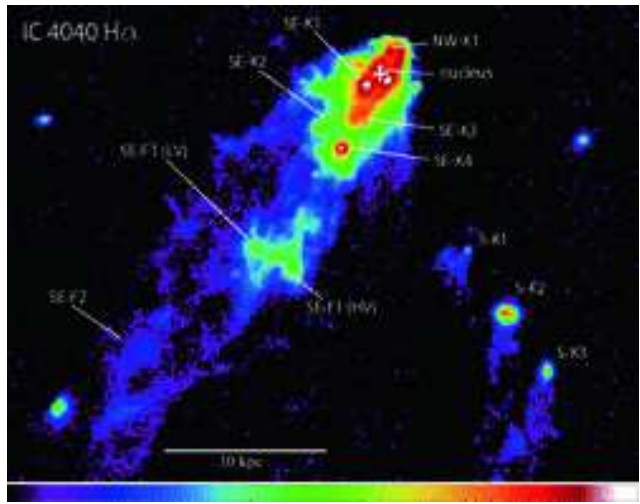}
\end{center}
\caption{
\label{IC4040expand} Details of the morphology of the H$\alpha$ emitting gas of IC~4040 around the galaxy disk
and nucleus. 
In the central region of the galaxy, several bright H$\alpha$ knots and complicated structure of 
ionized gas filaments/clouds can be seen. 
The H$\alpha$ gas is truncated at the north-western side and exhibits bow shock like morphology.
}
\end{figure}

\begin{figure}
\begin{center}
\includegraphics[scale=0.9]{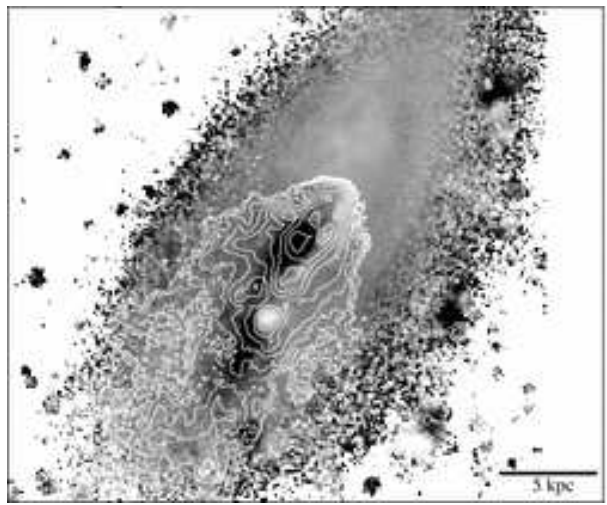}
\end{center}
\caption{
\label{IC4040B-R} $B-R_{\rm C}$ color image (gray-scale) of IC4040. 
The darker color represents a larger $B-R_{\rm C}$.
Contour map of the H$\alpha$ gas distribution is overlaid on the $B-R_{\rm C}$ image.  
}
\end{figure}

The $B - R_{\rm C}$ color image shown in Figure 9 reveals that the southeast
stream is very dusty.
The close spatial correlation between the dust and the EIG and the one-sided
morphology of the dust suggest that the ionized gas and the dust are 
stripped together and are mixed with each other in the stripped flow.
Both the dust stream and the southeast part of the EIG extend in the same
direction as the \ion{H}{1} gas tail found by \citet{Bravo2000,Bravo2001}.
Recent far-infrared/submm survey of the Virgo cluster with {\it Herschel} revealed
that dust disks are truncated in H~{\sc i} deficient galaxies \citep{Cortese2010}.
These results suggest that dust in galaxies can be stripped in the cluster environment.  
The dust stream in IC~4040 exemplifies a dust stripping site observed in a cluster.

The EIG is widely distributed south of IC~4040.
The southern part of the EIG consists of three conspicuous bright knots 
(S-K1, S-K2 and S-K3; see Figure 2), 
several giant complexes of clouds/filaments 
(S-F1, S-F2, etc.) and many small filaments.

The three bright knots are blue in color.
These features have been detected in the UV light with {\it GALEX} \citep{Smith2010}.
Of the three knots, knot 2 is the brightest.
\citet{Smith2010} observed a Wolf-Rayet feature in their Keck spectrum, 
suggesting this knot is very young (younger than 5 Myr).
Elongated H$\alpha$ filaments are connected to these knots.
All the connected filaments have head-tail morphologies whose position 
angles are the same as the southeast stream. 

Many ionized gas clouds/filaments are distributed over 30 kpc $\times$ 80 kpc 
in the south area of IC~4040.
The brightest one is a thick loop-like filament at about 40 kpc from the galaxy
(S-F1 in Figure 2). 
S-F2 is another bright filament connected to S-F1.
S-F3 and S-F4 are diffuse filaments seen near an early type galaxy GMP~2551
(Figure 7).
Other than these filaments, many small filaments are seen between
S-F2 and IC~4040.

\begin{figure}[htbp]
\begin{center}
\includegraphics[scale=0.33]{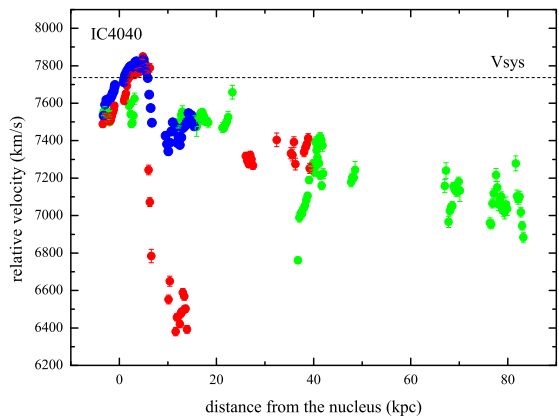}
\end{center}
\caption{
\label{IC4040vel} The velocity field of the EIG of IC~4040.
Green dots, blue dots and red dots represent the data points of the slit masks
of MS0631, MS0632, and MS0632-off, respectively. 
}
\end{figure}

Figure 10 shows the velocity field of the EIG of IC~4040.
The brightness of the EIG of IC~4040 allowed us to trace the velocity 
field along each slit.
The IC~4040 EIG has a velocity field similar to that of the RB199 EIG;
the relative radial velocities of the filaments increase almost 
monotonically with the distance from the nucleus (Figure 10).
Within 5 kpc of the nucleus, the H$\alpha$ gas along the galaxy major
axis follows the galaxy rotation.
The acceleration of the EIG begins at 5 kpc from the nucleus.
The overall acceleration rate in a range of 5 -- 80 kpc is 
$\sim -10$ km~s$^{-1}$~kpc$^{-1}$.

\begin{figure}
\begin{center}
\includegraphics[scale=0.33]{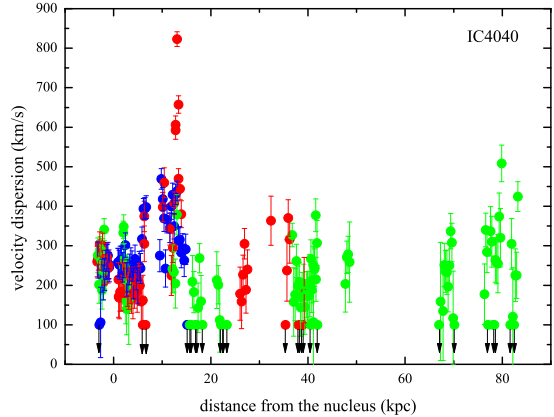}
\end{center}
\caption{
\label{IC4040veldisp} The velocity dispersion of the EIG of IC~4040.
The color difference is the same as Figure 10.
}
\end{figure}

We noted several peculiar features in the velocity field of the IC~4040 EIG.
The most remarkable one is a very low-velocity component at $\sim 12$
~kpc from the nucleus.
This component is spatially associated with the SE-F1 (see Figure 2); we call 
it SE-F1 (LV).
The relative velocity of the SE-F1 (LV) is $-1300$ km~s$^{-1}$ with 
respect to the systemic velocity of the galaxy.
Its velocity dispersion reaches  $\approx 800$ km~s$^{-1}$
(Figure 11).
In contrast, the southwest side of the SE-F1 is much faster than 
the SE-F1 (LV) and almost follows the overall motion of the EIG.
This fast component, called SE-F1 (HV) has a relative velocity of 
$\approx -300$ km~s$^{-1}$ and a velocity dispersion of $\approx 400$
km~s$^{-1}$.
Figure 12 shows the two-dimensional spectra of the circumnuclear 
H$\alpha$ gas and the southeast filaments.
In this figure, large velocity shift and wide velocity dispersion
of SE-F1 (LV) are clearly seen.

\begin{figure}[htbp]
\begin{center}
\includegraphics[scale=0.6]{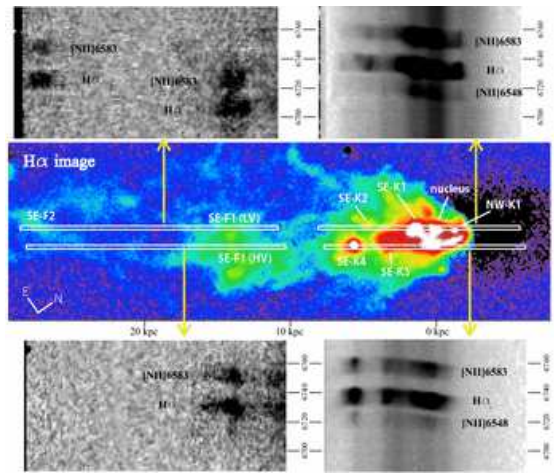}
\end{center}
\caption{
\label{IC4040Havel} Two dimensional spectra of the SE filaments and the nuclear H~{\sc ii} regions of IC~4040.
Middle panel shows a pseudo color image of the H$\alpha$ emitting filaments. 
}
\end{figure}

The S-F1 has a large velocity gradient ($\sim 70$ km~s$^{-1}$~kpc$^{-1}$)
within it.
The filament has a relative velocity of $\approx -700$ km~s$^{-1}$
at the edge of the galaxy.
At the opposite edge, the relative velocity reaches $-450$ km~s$^{-1}$.
The S-F3 and S-F4 are highly turbulent and the velocity dispersions
are $\sim 200 - 400$ km~s$^{-1}$.

\subsubsection{GMP~2923 and GMP~3071}

The EIG around GMP 2923 is shown in Figure 4d in YY10. 
Two galaxies --- GMP 2923 and GMP 2945 --- overlap in the line of sight
near the EIG (Figure 3). 
We measured the radial velocity of the early-type
galaxy GMP 2945 with the same slit mask as that used for the EIG;
it is 6500 km~s$^{-1}$, much slower than those of the EIG and GMP 2923.
Thus GMP 2945 is kinematically independent from the EIG.

The EIG of GMP~2923 is characterized by two knotty straight streams (see Figure 4d 
in YY10).
The opening angle of the streams is $\approx$20 degrees.
The root of the streams is detached from the galaxy.
The EIG is extended to $\sim 50$ kpc from GMP~2923. 

The EIG of GMP~3071 consists of slightly wiggled narrow filaments 
(Figure 4f in YY10). 
The root of the EIG is connected to the nucleus of the galaxy.
The angle between the galaxy major axis and the EIG is $\approx$40 
degrees and the length of the EIG is $\sim 25$ kpc.

\begin{figure}[htbp]
\begin{center}
\includegraphics[scale=0.33]{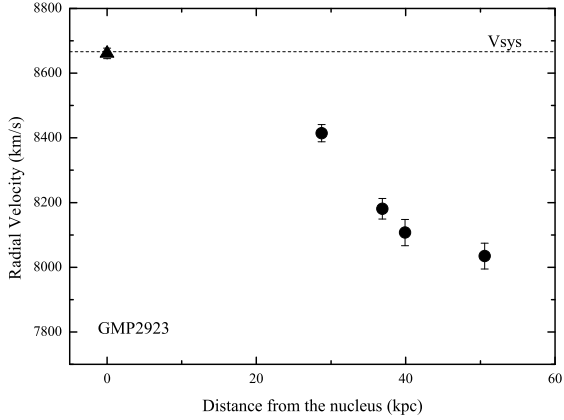}
\end{center}
\caption{
\label{GMP2923vel} The velocity field of the EIG of GMP~2923.
}
\end{figure}

\begin{figure}[htbp]
\begin{center}
\includegraphics[scale=0.33]{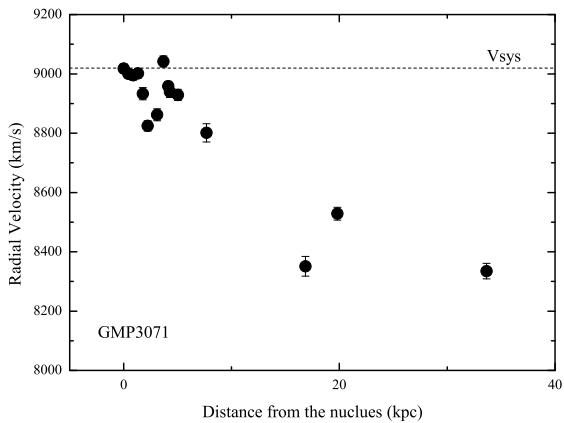}
\end{center}
\caption{
\label{GMP3071vel} The velocity field of the EIG of GMP~3071.
}
\end{figure}

The kinematics of both the GMP~2923 and GMP~3071 EIGs also reveal
a monotonically-increasing pattern.
Figure 13 and 14 show the velocity fields of the EIGs of the two galaxies.
The acceleration rates are $\approx -10$ km~s$^{-1}$~kpc$^{-1}$ and 
$\approx -25$ km~s$^{-1}$~kpc$^{-1}$
for GMP~2923 and GMP~3071, respectively.

The kinematics of the EIG of GMP~3071 is rather complex in the 
galaxy disk, showing a rapid drop-and-rise whose amplitude is 
$\approx 200$ km~s$^{-1}$ (Figure 14).
The net H$\alpha$ image taken by YY10 revealed that the
ionized gas near the nucleus has a wiggled chain of small blobs.
Together with the kinematics, this may suggest a helical motion of
the ionized gas caused by the stripping force coupled with galactic
rotation.

\subsection{Excitation of the EIGs}

Most of the EIG extended filaments are too faint to detect emission
lines other than H$\alpha$.
We could detect diagnostic emission lines such as [S~{\sc ii}] or 
[N~{\sc ii}] only in bright knots and some bright filaments in the EIGs
of RB199 and IC~4040.
Table 2 summarizes the measured fluxes of the emission lines for 
bright parts of the EIGs of RB199 and IC~4040.

\begin{table*}[tbp]
\caption{Emission Line Fluxes and Intensity Ratios of the EIGs.}
\begin{center}
\begin{tabular}{lcccccccc}
\hline
ID & ditance$^{a}$ & length$^{b}$ & $f_{{\rm H}\alpha}^{c}$ & $f_{{\rm H}\beta}^{d}$ & [NII]/H$\alpha^{e}$ &
[SII]/H$\alpha^{e}$ & [OI]/H$\alpha^{e}$ & [OIII]/H$\beta^{e}$ \\
\hline
{\bf IC~4040} & & & & & & & \\
nucleus  & $-$  & 0.59 & 271 & 85.3 & $-0.47\pm0.01$ & $-0.52\pm0.01$ & $-1.60\pm0.01$ & $-0.36\pm0.01$ \\
NW-K1    & 1.2  & 0.44 & 221 & 51.2 & $-0.48\pm0.01$ & $-0.61\pm0.01$ & $-1.69\pm0.02$ & $-0.54\pm0.01$ \\
SE-K1    & 1.1  & 0.74 & 303 & 53.6 & $-0.32\pm0.01$ & $-0.57\pm0.01$ & $-1.36\pm0.01$	& $-0.43\pm0.01$ \\
SE-K2    & 3.8  & 0.90 & 47.1 & 17.1 & $-0.40\pm0.01$ & $-0.44\pm0.01$ & $-1.22\pm0.03$ & $-0.51\pm0.03$ \\
SE-K3    & 2.5  & 0.74 & 107 & 23.9  & $-0.43\pm0.01$ &	$-0.58\pm0.01$ & $-1.42\pm0.02$ & $-0.27\pm0.01$ \\
SE-K4    & 4.8  & 0.74 & 194 & 37.1  & $-0.51\pm0.01$ &	$-0.61\pm0.01$ & $-1.61\pm0.02$ & $-0.17\pm0.01$ \\
S-K1     & 12.3 & 0.59 & 2.1 & $-$ & $-0.49\pm0.10$ &	$-0.12\pm0.07$ & $-0.48\pm0.10$ & $-$            \\
S-K2     & 17.0 & 0.74 & 172 & 59.1 & $-1.04\pm0.01$ &	$-1.05\pm0.01$ & $-1.98\pm0.04$ & $ 0.50\pm0.01$ \\
S-K3     & 21.7 & 0.74 & 14.6 & 5.3 & $-0.37\pm0.02$ & $-0.40\pm0.02$ & $-1.41\pm0.12$ & $-0.05\pm0.04$ \\
S-K4     & 47.9 & 0.74 & 3.2 & $-$ & $-$            & $-$            & $-$            & $-$            \\
SE-F1(LV)& 12.8 & 2.22 & 16.4 & 3.9 & $-0.12\pm0.03$ & $-0.51\pm0.09$ & $-0.64\pm0.09$ & $ 0.04\pm0.13$ \\
SE-F1(HV)& 13.8 & 3.11 & 28.4 & 3.4 & $-0.16\pm0.03$ & $-0.38\pm0.06$ & $-0.63\pm0.06$ & $ 0.12\pm0.21$ \\
SE-F2    & 26.7 & 1.78 & 14.8 & $-$ & $-$            & $-$            & $-$            & $-$            \\
S-F1N    & 37.8 & 0.59 & 5.7 & 2.1  & $-0.44\pm0.04$ & $-0.27\pm0.04$ & $-0.77\pm0.07$ & $-0.42\pm0.15$ \\
S-F1S    & 39.2 & 0.74 & 5.5 & 2.1  & $-0.47\pm0.05$ & $-0.42\pm0.06$ & $-0.90\pm0.11$ & $-0.36\pm0.18$ \\
S-F2     & 40.9 & 1.04 & 2.8 & $-$ & $-0.25\pm0.08$ & $-0.27\pm0.10$ & $-$            & $-$            \\
S-F3     & 68.8 & 2.37 & 6.3 & $-$ & $-1.08\pm0.22$ & $-$            & $-$            & $-$            \\
        & & & & & & & \\
{\bf RB199}   & & & & & & & \\
knot1    & 22.8 & 1.18 & 23.5 & 8   & $-1.02\pm0.03$ & $-0.67\pm0.02$ & $-1.52\pm0.11$ & $ 0.19\pm0.02$ \\
knot1e   & 23.7 & 1.04 & 3.8  & 0.7 & $-1.06\pm0.33$ & $-0.24\pm0.09$ & $-0.61\pm0.14$ & $-0.25\pm0.44$ \\
knot2    & 24.1 & 1.48 & 37.9 & 13  & $-0.86\pm0.02$ & $-0.29\pm0.01$ & $-1.06\pm0.03$ & $ 0.07\pm0.02$ \\
knot3    & 34.8 & 0.89 & 4.9  & 2   & $-0.92\pm0.09$ & $-0.26\pm0.03$ & $-0.99\pm0.10$ & $-0.22\pm0.11$ \\
knot4    & 36.2 & 0.74 & 1.9  & $-$ & $-$            & $-$            & $-$            & $-$            \\
knot5    & 34.9 & 1.04 & 21.6 & 8   & $-1.14\pm0.06$ & $-0.65\pm0.03$ & $-1.45\pm0.12$ & $ 0.37\pm0.02$ \\
H$\alpha$ filament 1(N) & 43.2 & 2.51 & 1.3 & $-$ & $-$ & $-$         & $-$            & $-$            \\
H$\alpha$ filament 1(S) & 50.4 & 1.18 & 2.4 & $-$ & $-$ & $-$         & $-$            & $-$            \\
H$\alpha$ filament 2    & 39.8 & 0.89 & 2.7 & $-$ & $-$ & $-$         & $-$            & $-$            \\
H$\alpha$ cloud 2       & 76.3 & 1.48 & 1.9 & $-$ & $-$ & $-$         & $-$            & $-$            \\
\hline
\end{tabular}
\end{center}

\footnotesize{
\noindent
$^{(a)}$ Distance from the nucleus of the parent galaxy in unit of kpc.\\
$^{(b)}$ Length along the slit in unit of kpc.\\ 
$^{(c)}$ H$\alpha$ flux in unit of $10^{-17}$ ergs s$^{-1}$ cm$^{-2}$.\\
$^{(d)}$ H$\beta$ flux in unit of $10^{-17}$ ergs s$^{-1}$ cm$^{-2}$.\\
$^{(e)}$ Logarithmic value of the emission line intensity ratio.
} 
\end{table*}

\begin{figure}[htbp]
\begin{center}
\includegraphics[scale=.44]{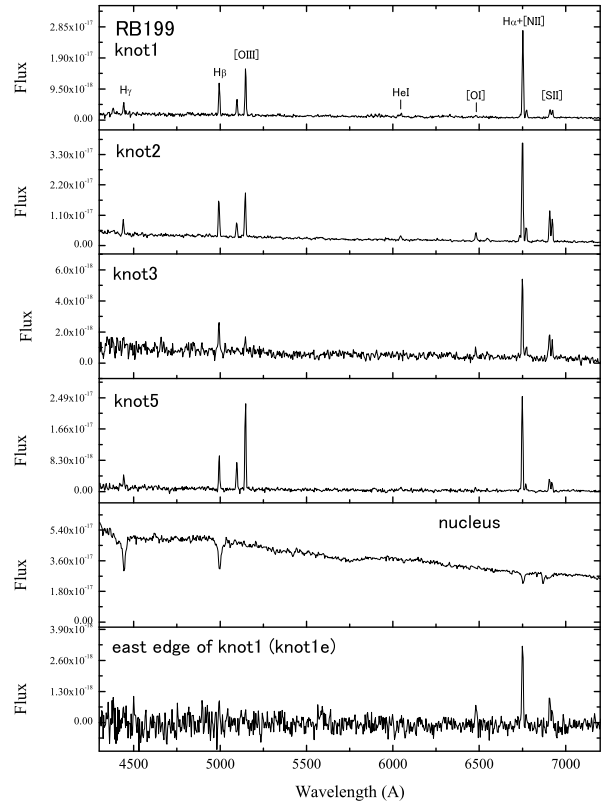}
\end{center}
\caption{
\label{RB199sp} Spectra of the bright knots of the EIG 
and the nucleus of RB199.
}
\end{figure}

Figure 15 shows the spectra of the bright knots and the nucleus of RB199.
The bright knots of RB199 are characterized by strong emission lines
and blue continuum.
The equivalent widths of H$\alpha$ are 310 \AA, 230 \AA, and 
1500 \AA\ for knot 1, 2, and 5, respectively.
The most interesting feature of the bright knots spectra is 
the weak [N~{\sc ii}]6548/6583 emission.
In knot 2 or knot 3, [S~{\sc ii}]6717/6731 lines 
are stronger than [N~{\sc ii}].
In knot 3, even [O~{\sc i}]6300 is stronger than [N~{\sc ii}].
In knot 1 and knot 5, very weak [N~{\sc ii}], [S~{\sc ii}] and [O~{\sc i}]
and relatively strong [O~{\sc iii}]5007 indicate that these 
knots are metal-deficient H~{\sc ii} regions.
The spectra of knot 2 and knot 3 reveal strong [S~{\sc ii}] and 
[O~{\sc i}] lines, suggesting that shock heating contributes the 
gas excitation.
The nucleus of RB199 has a spectrum typical of post-starburst galaxies.
It is characterized by strong Balmer absorption lines, very weak 
metal-absorption lines, and a blue continuum.

\begin{figure}
\begin{center}
\includegraphics[scale=.44]{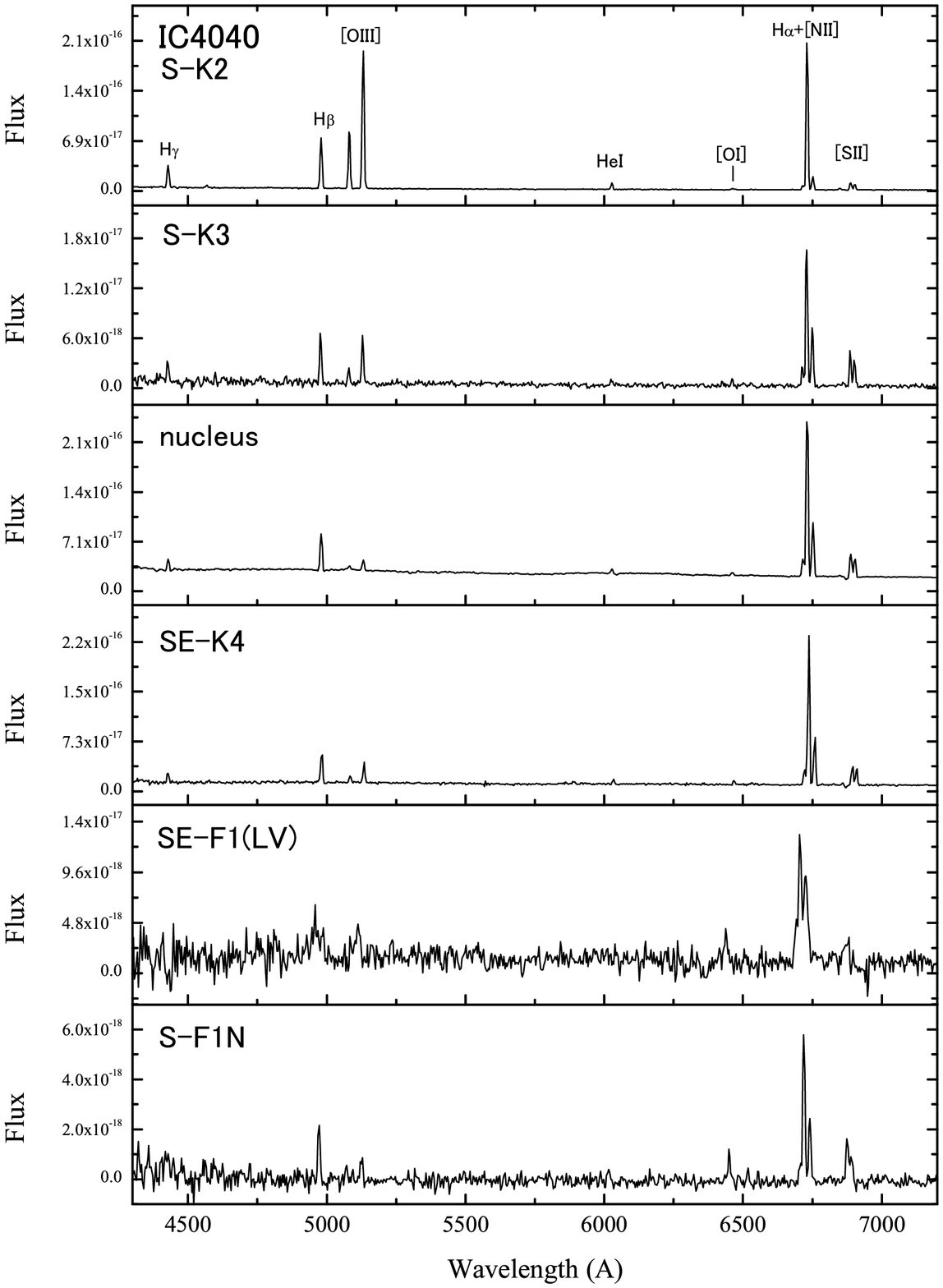}
\end{center}
\caption{
\label{IC4040sp} Spectra of the bright knots and filaments of the EIG 
and the nucleus of IC~4040.
}
\end{figure}

Figure 16 shows the spectra of the bright knots and filaments of the
EIG of IC~4040 with the nuclear spectrum of IC~4040 itself.
The spectrum of S-K2 of the IC~4040 EIG is that of a typical metal-poor
star-forming region.
The H$\alpha$ line is very strong in S-K2; the equivalent width reaches
$\sim 1000$ \AA.
In the blue region of the spectrum, we confirmed a weak Wolf-Rayet feature
(Figure 17) which was found by \citet{Smith2010}.
S-K3 shows relatively strong [N~{\sc ii}] and [S~{\sc ii}] suggesting shock
excitation may play a role in the gas excitation.
The spectra of the nucleus and S-K4 resemble each other, while
the continuum of S-K1 is redder than that of the nucleus.
It reflects the dusty gas flow in the disk of the galaxy (Figure 9) 
and S-K4 is embedded in the dusty flow.

\begin{figure}[htbp]
\begin{center}
\includegraphics[scale=.3]{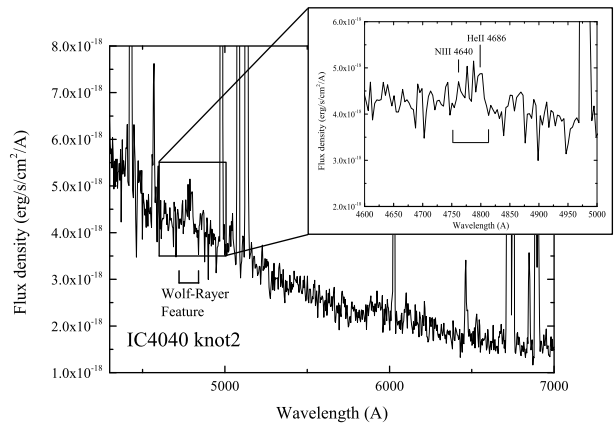}
\end{center}
\caption{
\label{IC4040WR} Expanded plot of the spectrum of S-K2 of the IC~4040 EIG.
A broad bump at around $\lambda$4650 \AA\ (Wolf-Rayet feature) can clearly be seen
(the significance of the detection is $\sim 3 \sigma$).
The insertion is an enlarged spectrum around the Wolf-Rayet feature.
Although signal-to-noise ratio is not sufficiently high to allow us to
decompose the feature into individual emission lines, 
HeII $\lambda$4686 \AA\ and NIII $\lambda$4640 \AA\ can be identified.
}
\end{figure}

Certain low ionization forbidden lines, [S~{\sc ii}], [O~{\sc i}] and 
[N~{\sc ii}], are enhanced relative to H$\alpha$ in some extended 
filaments of the IC~4040 EIG: SE-F1, S-F1 and S-F3 (Figure 16).
An extended filament attached to the tip of knot 1 
(east edge of knot 1) of RB199 also 
exhibits enhanced [S~{\sc ii}] and [O~{\sc i}], as shown in the bottom
panel of Figure 15.
Other extended filaments are too faint to measure the intensities of 
emission lines other than H$\alpha$.
%It is also noticeable that the emission line widths are broad in SE-F1 (LV);
%see Figure 16.

\begin{figure}[htbp]
\begin{center}
\includegraphics[scale=.33]{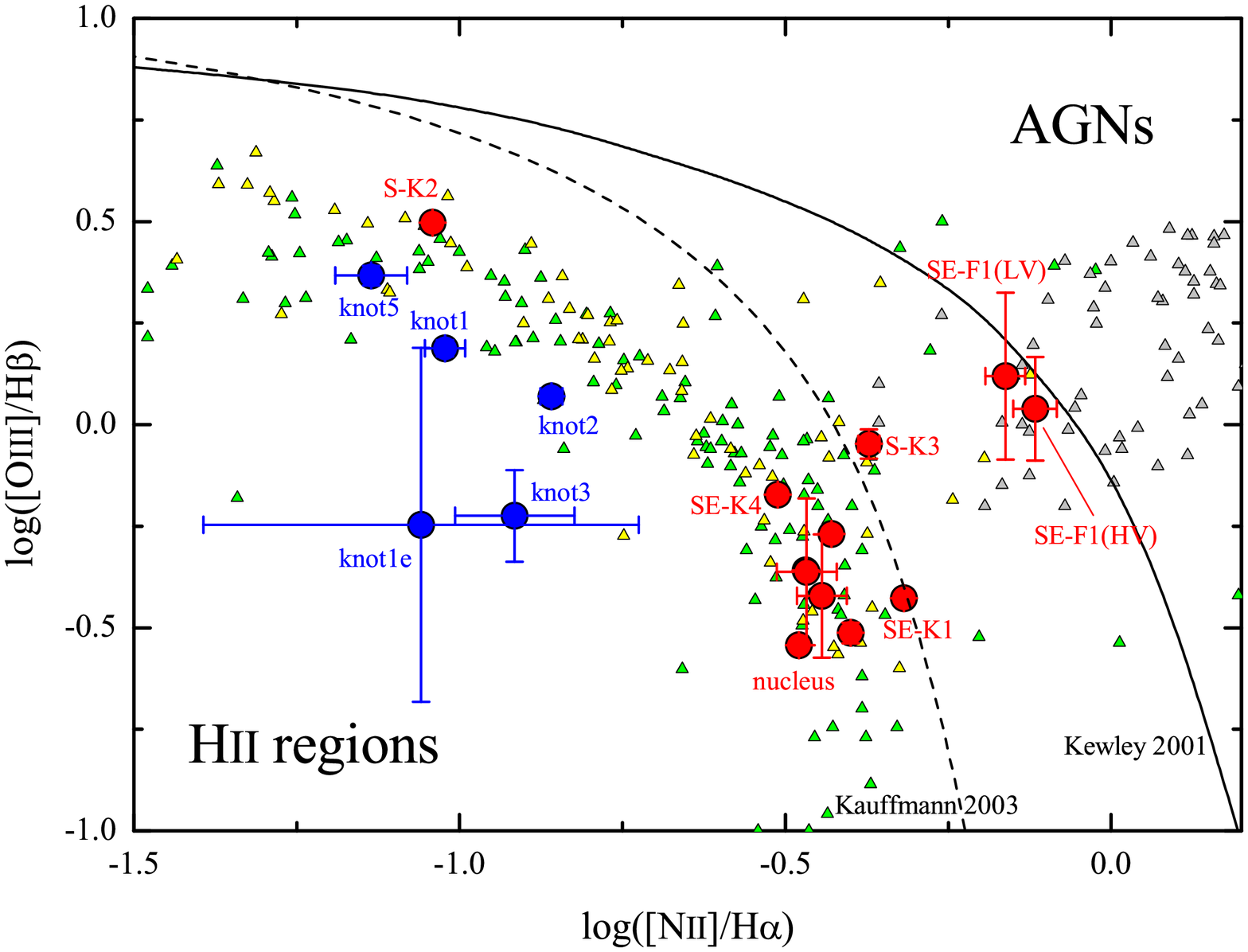}
\end{center}
\caption{
\label{N2O3} [O~{\sc iii}]5007/H$\beta$ ratio vs.
[N~{\sc ii}]6584/H$\alpha$
ratio diagram. 
Blue circles represent the data points of the bright knots of
the RB199 EIG (see Figure 1). Red circles represent the data of the
bright knots and filaments of the IC~4040 EIG (see Figure 2).
The solid curve and dashed curve are demarcations between H~{\sc ii} regions and
active galactic nuclei (AGN) proposed by \citet{Kewley2001} and \citet{Kauffmann2003},
respectively.
The green and yellow triangles are the data of H~{\sc ii} regions of nearby field galaxies
\citep{Jansen2000} and the data of blue compact galaxies \citep{Kong2002},
respectively.
The gray triangles are the data of LINERs \citep{Ho1997}.
}
\end{figure}

\begin{figure}
\begin{center}
\includegraphics[scale=.33]{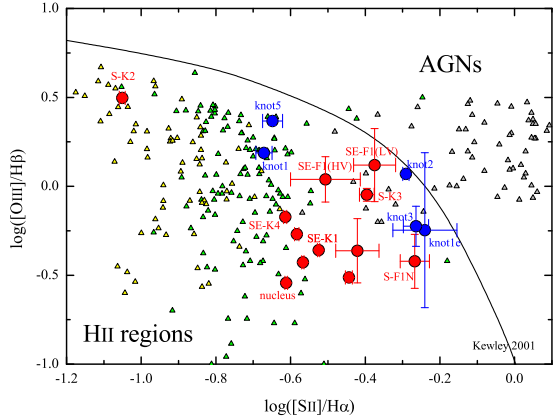}
\end{center}
\caption{
\label{S2O3} [O~{\sc iii}]5007/H$\beta$ ratio vs.
[S~{\sc ii}]6716+6731/H$\alpha$
ratio diagram.
Symbols and curves are the same as Figure 18.
}
\end{figure}

\begin{figure}[htbp]
\begin{center}
\includegraphics[scale=.33]{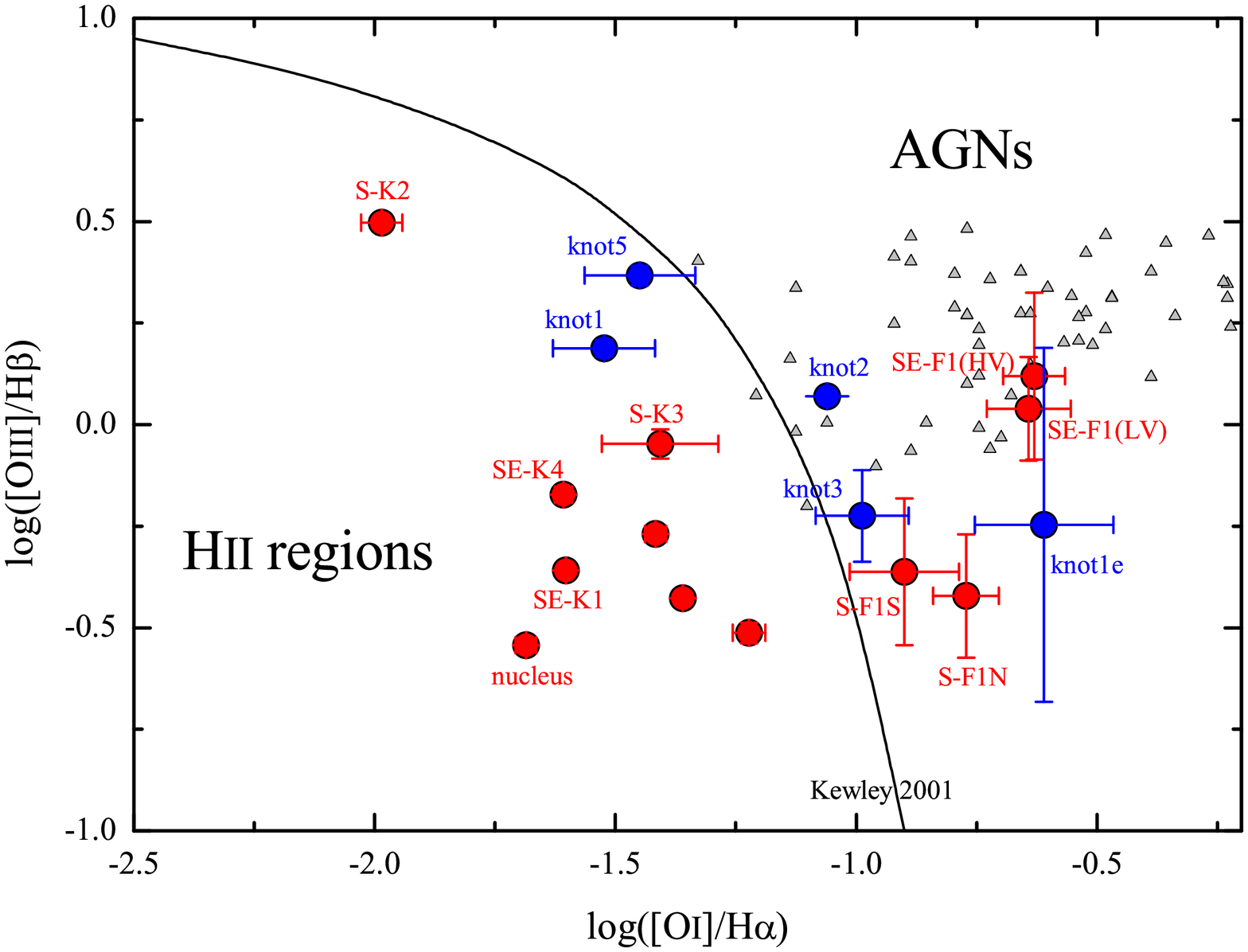}
\end{center}
\caption{
\label{O1O3} [O~{\sc iii}]5007/H$\beta$ ratio vs.
[O~{\sc i}]6300/H$\alpha$
ratio diagram.
Symbols and curves are the same as Figure 18.
}
\end{figure}

Figures 18, 19 and 20 are emission line diagnostic diagrams 
\citep{Veilleux1987} of the EIG knots and filaments of RB199 and IC~4040.
These diagrams clearly show the above-mentioned characteristics of the 
EIG spectra.
The relative weakness of [N~{\sc ii}]6583 in the RB199 EIG is conspicuous
in Figure 18; the emission line intensity ratio log([N~{\sc ii}]/H$\alpha$)
$\leq -1.0$ for the knots of the RB199 EIG.
Extragalactic H~{\sc ii} regions, which have low [N~{\sc ii}]/H$\alpha$ 
ratios, also generally have low [S~{\sc ii}]/H$\alpha$ and 
[O~{\sc i}]/H$\alpha$ ratios.
Figure 21 shows the abnormally weak [N~{\sc ii}] emission of 
the RB199 knots.
The data points of knots 1e, 2 and 3 are well displaced from the main
sequence of star-forming galaxies, whereas knots 1 and 5 are embedded 
in the sequence.
%Note that the region in which knots 1e, 2 and 3 are located is blank
%in Figure 21.

\begin{figure}[htbp]
\begin{center}
\includegraphics[scale=.33]{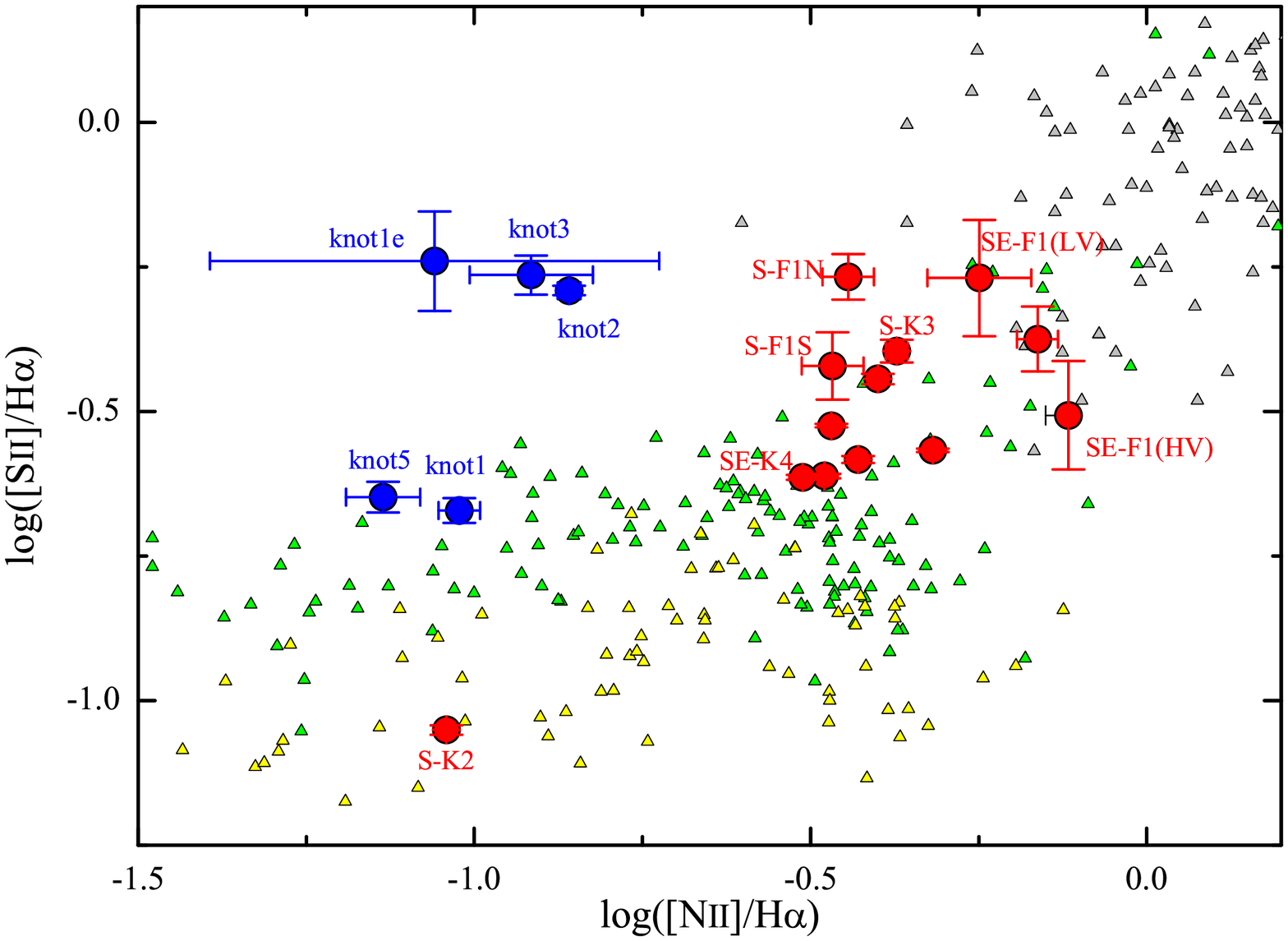}
\end{center}
\caption{
\label{N2S2} [N~{\sc ii}]6583/H$\alpha$ ratio vs.
[S~{\sc ii}]6716+6731/H$\alpha$
ratio diagram. Symbols are the same as Figure 18.
}
\end{figure}

The emission line ratios of the bright knots and the nucleus of IC~4040 
with the exception of S-K1 are consistent with those of star-forming 
galaxies (Figures 18, 19, 20 and 21).
Among these, it is clear that S-K2 of IC~4040 shows an emission 
line spectrum like blue compact dwarf galaxies.
In contrast, some extended filaments have LINER-like spectra.
In particular, the [O~{\sc i}] line is enhanced in these features 
(Figure 20). 
This [O~{\sc i}] enhancement suggests that shock heating plays an 
important role in ionization and excitation of these features.
The [N~{\sc ii}] weak features in the RB199 EIG also reveal similar 
[O~{\sc i}] enhancement, suggesting that shock heating is important 
for these peculiar features.
However, the abnormally weak [N~{\sc ii}] line of these features cannot
be explained by ordinary shock heating models.
Low velocity shock heated gas which enhances [O~{\sc i}] and [S~{\sc ii}]
overlapped on low metal abundance H~{\sc ii} region might produce
such peculiar emission-line intensity ratios.

\begin{table*}[tbp]
\caption{Kinematics and Abundances of the Bright Knots.}
\begin{center}
\begin{tabular}{lcccc}
\hline
ID & $v_{\rm rel}$ (km~s$^{-1}$) & $\sigma^{a}$ (km~s$^{-1}$) & log(O/H)+12 & $Z_{\odot}^{b}$ \\
\hline
    {\bf RB199}     & &  & \\
    knot1    & $-104\pm16$ & $< 70$ & $8.38\pm0.05$ & $0.49\pm0.08$ \\
    knot2    & $-194\pm15$ & $< 70$ & $8.48\pm0.05$ & $0.62\pm0.10$ \\
    knot3    & $-210\pm17$ & $< 70$ & $8.40\pm0.03$ & $0.51\pm0.07$ \\
    knot5    & $-240\pm15$ & $< 70$ & $8.44\pm0.05$ & $0.56\pm0.09$ \\
                    & &  & \\
    {\bf IC~4040}   & &  & \\
    S-K2     & $-241\pm18$ & $93\pm20$ & $8.59\pm0.03$ & $0.79\pm0.10$ \\
    S-K3     & $-268\pm16$ & $< 70$ & $8.88\pm0.05$ & $1.55\pm0.25$ \\
    SE-K4    & $-210\pm17$ & $< 70$ & $8.40\pm0.02$ & $0.51\pm0.06$ \\
\hline
\end{tabular}
\end{center}

\footnotesize{
\noindent
$^{(a)}$ Velocity dispersion $\sigma$ of H$\alpha$ line derived by gaussian fitting 
 ($f(\lambda) \propto {\rm exp}(-\lambda^2 / 2 \sigma^2)$).\\
$^{(b)}$ Oxygen abundances relative to the solar value derived by \citet{Asplund2009}
 (log(O/H)+12 $=$ 8.69$\pm$0.05).
}
\end{table*}

We tried to estimate the metal abundance of the bright knots around
RB199 and IC~4040 using the model calculations by \citet{Kewley2002}.
We used [N~{\sc ii}]/[O~{\sc iii}] and [N~{\sc ii}]/H$\alpha$ ratios
to estimate the metal abundance. 
The kinematical characteristics and oxygen abundances of the bright knots
are shown in Table 3.
The emission-line widths of most of the bright knots except for S-K2 are too narrow
to resolve with our spectral resolution, indicating that the knots are much
colder than the extended filaments.
As suggested by the emission line diagnostic diagrams (Figures 18, 19 and 
20), the bright knots except for S-K3 and SE-K4 have basically low oxygen abundances;
log(O/H)+12 $\sim 8.4 - 8.6$. 
These values correspond to $0.5 - 0.8$ $Z_{\odot}$.
S-K3 of IC~4040 shows super solar abundance ($\sim 1.6$ $Z_\odot$).
Because this knot is located at marginal region between H~{\sc ii} regions and
LINERs in the diagnostic diagrams, shock heating may contribute enhancement of the
forbidden lines, in particular, the [N~{\sc ii}] line, of this knot.

\section{Discussion}

\subsection{Comparison with ram pressure stripping models}

The one-sidedness of the morphologies of the EIGs of the Coma cluster
strongly suggests RPS as the primary formation mechanism
of these features (YY08; YY10).
All four of the observed EIGs exhibit a monotonically-increasing pattern
in their velocity fields.
These types of velocity structures have been suggested in many numerical
simulations of RPS
\citep{Vollmer2003,Roediger2005,Roediger2008,Tonnesen2009,Tonnesen2010}.

Recently, \citet{Tonnesen2009,Tonnesen2010} performed three-dimensional 
hydrodynamical simulations of RPS incorporating radiative cooling.
They found that radiative cooling fragments the disk gas of 
ram-pressured galaxy.
This fragmentation forms holes in the disk and the holes accelerate the
gas stripping.
As a result, stripped gas flows straight behind the disk, such that the
width of the wake of the stripped gas becomes much narrower than the
width observed in models without cooling \citep{Tonnesen2009}.
\citet{Tonnesen2010} examined observable values such as the density, 
velocity and morphology of RPS wakes formed in simulations.

We compared the velocity fields of the EIGs around the Coma galaxies 
with those of the RPS simulations conducted by \citet{Tonnesen2010}.
Before making the comparison, we roughly estimated the age of the EIGs.
The velocity gradient of the EIGs of RB199 and IC~4040 is $\sim 7 - 10$
km~s$^{-1}$~kpc$^{-1}$.
The distances from the parent galaxies to the most distant filaments of
the EIGs are $\sim 80 - 90$ kpc.
Assuming uniform acceleration with distance, $a$ km~s$^{-1}$~kpc$^{-1}$, 
it takes $9.5\times10^{2}\cdot{\rm ln}(d) / |a|$ Myr for the filaments
to reach current locations, where $d$ is the distance from the galaxy in
the unit of kpc. 
In the case of the most distant filaments of RB199 and IC~4040,
it would thus be $\sim 700 - 900$ Myr.
Note that this number is the maximum value, because the real flow is
not a stationary accelerated one.
The most distant part of the flow would be formed by the gas
which was rapidly accelerated and reached its maximum speed in the early
stage of the stripping event.
Thus the minimum age of the wake would be simply given by the length divided by
the current velocity at the tip, $\sim 100$ Myr.
These values are consistent with those predicted for the 100-kpc scale
wakes formed in RPS simulations \citep{Abadi1999,Vollmer2001b,Schulz2001,
Roediger2005,Jachym2007,Roediger2008,Tonnesen2010}.

\begin{figure}[htbp]
\begin{center}
\includegraphics[scale=.35]{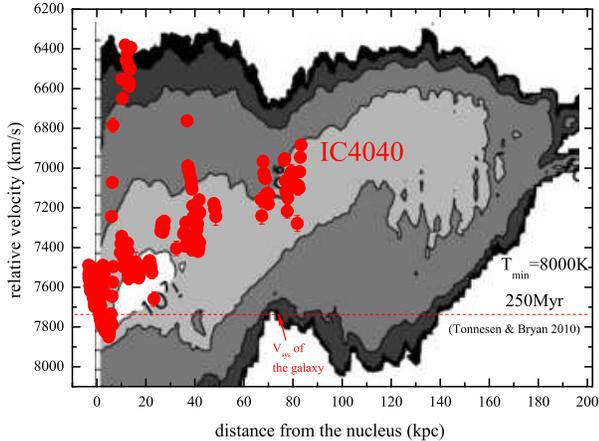}
\end{center}
\caption{
\label{IC4040vel-model} The velocity field of the EIG of IC~4040 (red circles) overlaid on that predicted
by a numerical simulation made by \citet{Tonnesen2010}.
The observational data are not corrected for projection effect.
The gray scaled contours represent the velocity field of the stripped gas in the simulation
at a time 250 Myr after the collision between the model galaxy and the ICM
\citep[reproduced by permission of the AAS]{Tonnesen2010}.
The collision velocity is 1413 km~s$^{-1}$. 
The cut off temperature of the cooling curve of the model is $T_{\rm min} = 8000$ K.
}
\end{figure}

In the simulations conducted by \citet{Tonnesen2009,Tonnesen2010}, the 
velocity field of the stripped gas is highly turbulent in the early 
phase of RPS.
Some gas components have very high velocities, close to the colliding 
velocity between the galaxy and the ICM, while some gas components have
negative relative velocities with respect to the galaxy.
After 250 Myr from the time of the galaxy--ICM collision, the velocity
field morphology begins to smooth out.
The relative velocity monotonically increases with the distance from the galaxy;
the main part of the stripped gas has a relative velocity of 900 
km~s$^{-1}$ at 150 kpc from the galaxy.
The width of the velocity distribution is about 700 km~s$^{-1}$.
Figure 22 overlays the data points for IC~4040 on a velocity field map taken
from \citet{Tonnesen2010}.
Clearly, the overall trend of the EIGs' velocity fields is consistent 
with the simulation.
The acceleration rates of the stripped gas are $7 - 10$ km~s$^{-1}$~kpc$^{-1}$
for RB199 and IC~4040, which is comparable to that of the main stream of the wake of the
simulation; $\approx 6$ km~s$^{-1}$~kpc$^{-1}$.
Additionally, the simulation predicts that part of the stripped gas near
the parent galaxy should have a very high relative velocity of 
$\sim 1000 - 1400$ km~s$^{-1}$ (Figure 22). 
This feature corresponds to SE-F1 (LV) of IC~4040 (SE-F1(LV)).

The main difference between the simulations and the observed data is 
the width of the velocity distribution.
Numerical simulations made so far have predicted wide velocity distribution
($\sim 1000$ km~s$^{-1}$) perpendicular to the stream direction.
\citet{Tonnesen2010} predicted much narrower velocity distribution 
(700 km~s$^{-1}$) by incorporating
radiative cooling, but it is still larger than that seen in the observations.
The velocity fields of the RB199 and IC~4040 EIGs are well-ordered 
except for SE-F1(LV).
The width of the velocity distribution of the EIG of IC~4040 is 
about $300 - 400$ km~s$^{-1}$,
which is a factor of two narrower than the simulations.
This discrepancy may be partly due to insufficient spatial coverage and
detection limit in our observations.
We observed only the bright parts of the EIG filaments and could not
follow the faint thin clouds, which may have much higher velocities
than the bright dense clouds.
Further, the spatial coverage of our observation was limited by slits.
Thus, our observation did not cover the entire velocity field of the EIGs.
A wide field, two dimensional spectroscopy would be needed for further
quantitative comparison between observations and the simulations.

The morphology of the wake of the stripped gas also differed between the 
simulations and the observation.
While many simulations predict a widely spread wake, most EIGs in the Coma
cluster have narrow, straight morphologies; the most remarkable example is the EIG
around D100 \citep{Yagi2007}.
The EIGs around GMP~2923 also has a narrow morphology.
The EIGs around GMP~3071 and RB199 are wider, but the widths do not exceed
the size of the galaxies.
Evaporation of outer thin component of the stripped gas or some confinement
mechanisms may narrow the morphology of the EIGs.

\subsection{Star formation in the EIG}

Strong star formation activities can be seen in the bright knots in the 
EIGs of RB199 and IC~4040. 
YY08 suggested, based on  $B$, H$\alpha$ and {\it GALEX} UV images,
that the bright knots in the EIG of RB199 are star-forming regions.
%S-K1, S-K2 and S-K3 of IC~4040 EIG are also bright in UV light \citep{Smith2010}.
\citet{Smith2010} found asymmetric UV emission for 13 galaxies in the Coma center.
They suggested that these emission regions were produced by active star 
formation in the stripped gas from their host galaxies.
The sample of \citet{Smith2010} largely overlaps  
the EIG sample presented by YY10.
Although YY10 did not identify optical continuum counterparts 
for most of the EIGs, \citet{Smith2010} pointed out that star formation 
activities are often associated with the EIGs (on the basis of their UV 
observations).
%One example is the 60 kpc H$\alpha$ tail associated with 
%D100 \citep{Yagi2007}.
%It is traced also by a deep $u$ band image taken with 
%CFHT, suggesting massive stars are formed in the tail \citep{Smith2010}. 

Recent numerical simulations of RPS have predicted active star formation
in the stripped gas \citep{Kronberger2008,Kapferer2008,Kapferer2009,
Tonnesen2010,Tonnesen2011}.
YY08 reported good morphological agreement between the
bright knots of the EIG of RB199 and the numerical simulation of 
\citet{Kronberger2008} (see also \citet{Kapferer2009}).
Although the spatial resolution of these simulations is insufficient 
to reproduce molecular gas formation in the stripped gas, 
\citet{Yamagami2011} recently found that molecular clouds can
be condensed in the RPS tails.
They suggested that magnetic fields would play an important role in
compressing the molecular clouds \citep{Yamagami2011}. 

\begin{figure}[htbp]
\begin{center}
\includegraphics[scale=.35]{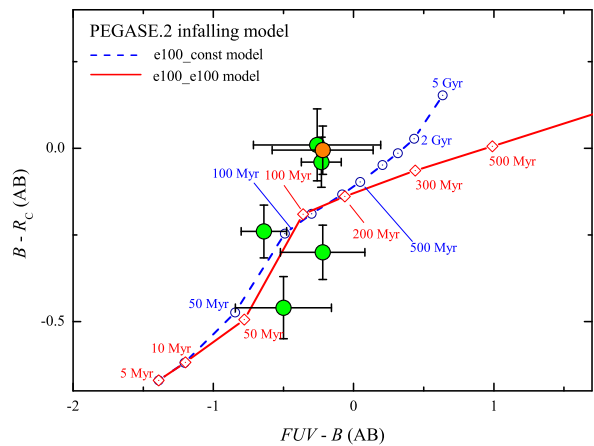}
\end{center}
\caption{
\label{RB199FUV-B} A $B - R_{\rm C}$ vs. $FUV - B$ color-color diagram of the bright knots of the
EIG of RB199 and IC~4040.
Green points and orange points represent the data of the RB199 knots and the IC~4040 knots, respectively.
Red line and blue dashed line show loci of star formation models
(using PEGASE.2 \citep{Fioc1997}, see the text).
Both models assume that stars are formed by infalling gas, and
the infall rate is assumed to be proportional to ${\rm exp}(-t / t_{\rm i}) / t_{\rm i}$, where
$t_{\rm i}$ is the timescale of the infall.
We adopted $t_{\rm i} = 100$ Myr.
Red line is the result of an exponentially decaying star formation model with a decay 
timescale of 100 Myr.
Blue dashed line is the result of a constant star formation model. 
}
\end{figure}
  
We compared the broad-band colors of the star-forming knots of 
RB199 (knot 1 to knot 5) and IC~4040 (S-K2) with star formation model.
Figure 23 is a $B - R_{\rm C}$ versus ${\rm FUV} - B$ color-color plot for
the star-forming knots.
The loci of star formation model calculations using PEGASE.2 
\citep{Fioc1997} are also plotted in the figure.
In these calculations, we assumed that stars are formed by infalling
of gas.
The infall rate was assumed to be proportional to 
${\rm exp}(-t / t_{\rm i}) / t_{\rm i}$,
where $t_{\rm i}$ is the infall timescale.
We adopted $t_{\rm i} = 100$ Myr.
We also assumed that the initial mass function (IMF) follows Salpeter IMF from 0.1 
M$_\odot$ to 120 M$_\odot$, and the abundance of the infall gas is 0.6 $Z_\odot$.
We did not include emission-line components in color calculation.
Figure 23 presents two cases; an exponentially-decaying star formation
model with a decay timescale of 100 Myr, and a constant star formation model. 
YY08 showed similar plots for the RB199 knots.
However, we found that the filter transmission functions of the $GALEX$
UV bands adopted by YY08 were incorrect, 
and the predicted ${\rm FUV} - B$ colors were significantly shifted to the blue direction.
Additionally, they did not subtract H$\alpha$ emission contribution 
from the $R_{\rm C}$ band data.
In this work, we used the correct $GALEX$ FUV band transmission curve
\citep{Morrissey2005} 
to calculate FUV magnitude of the models, and
the $R_{\rm C}$ magnitudes from which the H$\alpha$ emission flux
was carefully subtracted based on our spectroscopy.

Figure 23 shows that the star-forming knots are very young, on the
order of 100 Myr.
These bright knots are located at $\sim 20$ kpc from the parent galaxies
with a current relative velocity of $\sim 200$ km~s$^{-1}$.
Thus it would take at least $\sim 100$ Myr for these knots to reach the 
current positions.
This is comparable to or slightly shorter than the age estimation for 
the knots.
However, the real travel time of the knots
would be a few times longer than this value, because the knots are 
currently accelerated.
Together with the low metal abundance nature of the knots (see section 3.2),
this thus suggests that the most part of the stellar component of
these knots was formed in the stripped metal poor gas
which originally resided in the outskirt of the parent galaxies.
Current active star formation associated with the knots also supports
this idea.
Using hybrid N-body / hydrodynamical simulations, \citet{Kapferer2009}
found that 95\%\ of stars newly formed by RPS reside in the wake of 
stripped gas.
In their simulations, the timescale of star formation in the wakes is 
$\sim 100$ Myr \citep{Kapferer2009}.
The young age of the star-forming knots of the EIGs suggested by our 
results is consistent with their findings.

\subsection{Overlapped ionized gas in the EIG of RB199}

We found that an ionized gas cloud overlaps the line of sight to 
H$\alpha$ cloud 2 of RB199.
This cloud has a radial velocity of 6500 km~s$^{-1}$, which is
much slower than that of the EIG ($\sim 8000$ km~s$^{-1}$) and 
closer to the mean velocity of the Coma cluster.
It indicates that this overlapped cloud is kinematically independent
with the EIG.

What is this cloud and where did it originate?
One possibility is that it is a gas cloud stripped from a 
nearby galaxy.
A candidate of the parent galaxy is an early type galaxy 
(SDSS J125844.32+274251.6) near H$\alpha$ cloud 2.
According to the SDSS database\footnote{http://www.mpa-garching.mpg.de/SDSS/DR7/},
the color of this galaxy is that of a typical 
of S0 galaxies.
Unfortunately, the recession velocity of this galaxy is not known.
No morphological indication suggests a physical connection between 
this galaxy and the overlapped cloud.
We thus concluded that this galaxy is plausibly not the
parent of the cloud.

On the other hand, this cloud may be a remnant of a stripping 
event which occurred long ago.
In this case, identifying the parent galaxy would be difficult
because an apparent physical connection between them would be almost
completely lost.

Recently, a huge intra-cluster stellar population has been
found in local- and intermediate-redshift clusters
\citep{Gregg1998,Gonzalez2000,Mihos2005,Krick2007}.
Such population is thought to be originated by galaxy-galaxy interaction
or merger in the central region of the clusters.
Tidal force associated with galaxy-galaxy interaction would strip the 
stars from the galaxies and spread them into the intra-cluster space.
Associated with these tidal stripping events, dense disk gas may be also 
stripped, and a dense gas cloud may be floating in the 
intra-cluster space. 

Although it is impossible to reach any conclusions from this rare example,
this finding may suggest that there is a population of `floating ionized gas'
in clusters of galaxies \citep{Gerhard2002,Cortese2004}.
Deep H$\alpha$ imaging of intra-cluster space and spectroscopic follow up 
will reveal how common this kind of objects are in galaxy clusters.

\subsection{Fate of the EIG}

Our spectroscopic observations showed that it is plausible that
the EIGs associated to the four galaxies, RB199, IC~4040, GMP~2923
and GMP~3071, are formed by RPS.
The observations also made it clear that strong star formation occurs in 
some parts of the stripped gas.

The EIGs around galaxies in the Coma cluster may act as a snapshot of a
formation process of very faint galaxy population in clusters.
The typical brightness and size of the bright knots in the EIGs
are $M_R \sim -12 - -13$ and $1 - 2$ kpc (corresponding half light radius  
$\approx 200 - 300$ pc), respectively, which are similar to those of 
ultra compact dwarf galaxies \citep{Drinkwater2004} or small dwarf spheroidals.
This suggests that the bright knots are gravitationally self-bound systems
\footnote{Of course, measuring stellar velocity dispersion of the knots is necessary to
check whether they are really self bound systems or not, but the spectral
resolution and coverage of our spectroscopic data are not enough to 
perform such study.}.
In this case, they would survive and float in the halo of the parent
galaxies or in the intra-cluster space of the Coma.
After the star formation ceased, the luminosities of the bright knots 
will significantly decrease.
Assuming the exponentially decaying star-formation model we used in Section
5.2, and assuming a typical age for the bright knots is 200 Myr, we can
conclude that the luminosities of the knots would decrease by 
$2 - 3$ magnitudes in $R$ band in 1 Gyr.
That is, the absolute magnitudes of the knots will eventually be 
$M_R \sim -10 - -11$.
This luminosity range is within that of the dwarf spheroidal galaxies in
the local group \citep{Belokurov2007}.

Most of the stars of the EIGs may be bound to the
gravitational field of the parent galaxies.
\citet{Kronberger2008} discussed that most of the stars formed in the 
stripped gas in their simulations are gravitationally bound to the host 
galaxy and will fall back to the galaxy within 1 Gyr.
In this case, the bright knots may be observed as features like `galaxy aggregates' 
identified in the Coma cluster by \citet{Conselice1998} in the 
fall-back process. 
Because stellar systems are collision-less ones, the stars will penetrate
the galaxy disk and oscillate between two sides of the disk.
Accordingly, the bright knots and the blue filaments will 
eventually be destroyed and distributed widely in the galaxy halo;
in other words, the knots would form the halo population of the galaxy.

However, our spectroscopy revealed that the ionized gas
of the knots has the relative velocity of $200 - 270$ km~s$^{-1}$
(Table 3).
Because the stellar components of the knots are closely associated with
the ionized gas, the stars would have almost the same velocity as the 
ionized gas.
In this case, the knots probably escape from the gravitational
potential of the parent galaxies and float into the intra-cluster space.
In this context, it is interesting to note that an intra-cluster
H~{\sc ii} region was found in the Virgo cluster \citep{Gerhard2002}.

\section{Summary}

We have presented the results of deep imaging and spectroscopic observations of very 
extended ionized gas (EIG) around four member galaxies of the Coma cluster, 
RB199, IC~4040, GMP~2923 and GMP~3071.
Among them, the bright EIG around IC~4040 is remarkable and has complex morphology.
It consists of two streams extending towards the southeast and
southern directions from the galaxy.
The southeast stream is connected to the central active star formation region.
Many small ionized clouds are distributed between the southern stream and IC~4040.
Several star-forming blue knots are embedded in the EIGs of RB199 and IC~4040.
The extension of the EIGs ranges from $\sim 35$ kpc to $\sim 80$ kpc. 

The relative radial velocities of the EIGs with respect to the systemic
velocity of the parent galaxies from which they emanate increase
almost monotonically with the distance from the nucleus of the
respective galaxies, reaching $\sim -400 - -800$ km~s$^{-1}$ at
around $40 - 80$ kpc from the galaxies.
The velocity decreasing rates are $-7 - -10$ km~s$^{-1}$~kpc$^{-1}$ for
RB199, IC~4040 and GMP~2923.
The GMP~3079 EIG has the steepest velocity gradient of $-25$ km~s$^{-1}$~kpc$^{-1}$.
The kinematics of the EIGs is consistent with the concept that 
high-speed collisions between the intra-cluster medium and infalling galaxies
of the Coma cluster form these features, because all the sample galaxies have
systemic velocities much higher than the mean recession velocity of the cluster.  
We obtained a detailed velocity field for the bright EIG of IC~4040.
The velocity dispersions of the diffuse filaments of the IC~4040 EIG 
reach $\sim 200 - 400$ km~s$^{-1}$.
We also found a very low velocity filament at the southeastern edge of the disk of
IC~4040.
The filament has a velocity of $-1300$ km~s$^{-1}$ relative to the systemic velocity 
of the galaxy.

Some of the bright compact knots in the EIGs of RB199 and IC~4040 have blue continuum 
and strong H$\alpha$ emission.
The equivalent widths of the H$\alpha$ emission of the knots range $200 - 1500$ \AA.
The emission line intensity ratios of the knots are basically consistent with
those of sub-solar abundance \ion{H}{2} regions.
These facts indicate that
intensive star formation occurs in the knots.
Some filaments, including the low velocity filament of the IC~4040 EIG, exhibit
shock-like emission line spectra, suggesting that shock heating plays an 
important role in ionization and excitation of the EIGs.

The morphologies and velocity fields of the EIGs are broadly 
consistent with those predicted by numerical simulations of ram pressure stripping.
Our spectroscopic observations did not cover the entire regions of the EIGs
due to spatial limitation of slits.
This prevented us from making detailed comparison between the observation and 
the simulations.
Deep and wide integral field spectroscopy would allow us to compare 
the kinematics of the EIGs with the prediction of numerical simulations in detail.

\acknowledgments

We are grateful to the staff of the Subaru telescope for their kind
help with the observations.
We thank the anonymous referee for valuable comments which significantly 
improve this paper.
This work was in part carried out using the data obtained by a collaborative
study on the Coma cluster.
This study was in part carried out using the facilities at
the Astronomical Data Analysis Center (ADAC), National Astronomical
Observatory of Japan. This research made use of NASA's
Astrophysics Data System Abstract Service, NASA/IPAC Extragalactic
Database, SDSS skyserver, and GALEX GR5 database.
This work was financially
supported in part by Grant-in-Aid for Scientific Research 
No.23244030 from the Japan Society for the Promotion of Science
(JSPS) and No.19047003 from the Ministry of Education, Culture, Sports, 
Science and Technology (MEXT).

%----------------------------------------------------------------------------

%----------------------------------------------------------------------------
%\onecolumn

%-------------------------------------------------------------

%----------------------------------------------------------------------------

\end{document}